\documentstyle[multicol,aps,psfig]{revtex}
\renewcommand{\narrowtext}{\begin{multicols}{2}
\global\columnwidth20.5pc\noindent}
\renewcommand{\widetext}{\end{multicols}
\global\columnwidth42.5pc}
\begin{document}
\draft
\preprint{November 1999}
\title{Magnetic properties of frustrated spin ladder
}
\author{T\^oru Sakai and Nobuhisa Okazaki}
\address
{Faculty of Science, Himeji Institute of Technology,
 Ako, Hyogo 678-1297, Japan}
\date{November 1999}
\maketitle
\begin{abstract}
The magnetic properties of the antiferromagnetic spin ladder with 
the next-nearest neighbor interaction, particularly under external field, 
are investigated by the exact diagonalization of the finite clusters 
and size scaling techniques. 
It is found that there exist two phases, the rung-dimer and rung-triplet
phases, not only in the nonmagnetic ground state but also magnetized
one, where the phase boundary has a small magnetization dependence.    
Only in the former phase, the magnetization curve is revealed to have 
a possible plateau at half the saturation moment, with a sufficient 
frustration. 

\end{abstract}
\pacs{PACS numbers: 75.10.Jm, 75.30.Kz, 75.50.Ee, 75.60.Ej}
\narrowtext

\section{Introduction}
The frustration of the antiferromagnetic exchange interaction 
brings about many interesting phenomena in the quantum spin systems, 
because it generally enhances the quantum fluctuation. 
It would be valuable to consider the effect of the frustration 
on the spin ladder, like the materials 
SrCu$_2$O$_3$ (Ref.\cite{azuma}), Cu$_2$(C$_2$H$_{12}$N$_2$)$_2$Cl$_4$
(Refs.\cite{hayward,chaboussant}) and La$_6$Ca$_8$Cu$_{24}$O$_{41}$ 
(Ref.\cite{imai}).
They are strongly quantized and have the spin gap.  
When the next-nearest-neighbor
(NNN) exchange interaction appears,  
the frustration takes place in the system. 
In the classical limit it is easily shown that the system has two 
different ordered phases depending on the strength of the NNN exchange 
and the phase boundary does not change even under external magnetic 
field. 
In the quantum system, however, some modifications should exist  
in the ground state phase diagram, because the spin ladder has 
no long range order even at $T=0$.  
In this paper, 
we investigate the frustrated spin ladder by the exact diagonalization
of the finite clusters to determine the magnetic phase diagram, 
even under external field.  
In addition we consider the possibility of the magnetization plateau 
\cite{oshikawa}, 
which is predicted by a strong coupling approach.\cite{mila} 

\section{Model and numerical method}
The $S=1/2$ spin ladder with NNN coupling 
is described by the Hamiltonian
\begin{eqnarray}
\label{ham}
{\cal H}&=&J_1\sum_i^L({\bf S}_{1,i} \cdot {\bf S}_{1,i+1}
+{\bf S}_{2,i} \cdot {\bf S}_{2,i+1})\nonumber\\
 &+&J_{\perp}\sum_i^L({\bf S}_{1,i} \cdot {\bf S}_{2,i})\nonumber\\
 &+&J_2\sum_i^L({\bf S}_{1,i}
\cdot {\bf S}_{2,i+1}+{\bf S}_{2,i} \cdot {\bf S}_{1,i+1}), 
\end{eqnarray}
where $J_{1}$, $J_2$ and $J_{\perp}$ are the coupling constants
of the leg, NNN (diagonal) and rung exchange interactions,
respectively. We put $J_{\perp}$=1 in the following.
Using the Lanczos algorithm we numerically solved the ground state 
of the finite clusters. 
We also calculated the lowest energy
of ${\cal H}$ for $\sum_i^L({S_{1,i}^z}+{S_{2,i}^z})=M$
, which denotes $E(M)$. 
Using $E(M)$, we investigate the magnetic state with 
$m\equiv M/L$ under the external field
described by 
${\cal H}_Z=-H\sum_i^L({S_{1,i}^z}+{S_{2,i}^z})$.

\section{Two magnetic phases}
Consider the nonmagnetic ground state at first. 
In the classical limit the system has two different ordered phases 
divided by the first-order phase boundary $J_2=J_{\perp}/2 (=1/2)$ 
shown as a dashed line in Fig. 1. $J_1=1/2$ is also the boundary 
because the phase diagram should be symmetric
under the exchange of $J_1$ and $J_2$ (the reflection with respect to
the dot-dashed line in Fig. 1).  
\begin{figure}[htb]
\begin{center}
\mbox{\psfig{figure=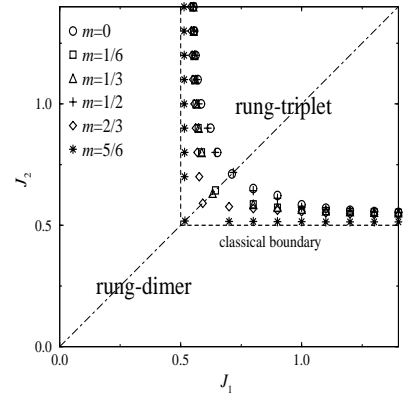,width=6cm,height=6cm,angle=-90}}
\end{center}
\caption{
Phase diagram of the frustrated spin ladder with magnetization
$m$ for $L=12$. Every phase boundary is first-order within this 
region. The dashed line is the classical limit. 
The dot-dashed line is the symmetric line. 
}
\label{fig1}
\end{figure}
In the quantum $S=1/2$ system we should distinguish the two phases 
based on the dimer picture; the dimers along the rung 
and the diagonal, respectively. The former is realized for $J_2 \ll
J_{\perp}/2$, while the latter for $J_2 \gg J_{\perp}/2$. 
In the latter phase each two spins coupled by the rung are expected to
behave like an effective $S=1$ (triplet) object. 
Thus we call the two phases `rung-dimer' and `rung-triplet',
respectively. 
The phase boundary is easily detected as a level crossing point in the
ground state even in small finite clusters. Since the boundary is almost 
independent of $L$, we show only the result of $L=12$ as circles in Fig.
1. 
Our study of the spin correlation function along the rung also 
supported the above argument and suggested that the boundary is 
first-order. 
The results are completely consistent with the recent analysis by the density 
matrix renormalization group. \cite{wang} (It also indicated the crossover 
of the phase boundary from first-order to second-order ones for 
$J_2 < 0.287 J_1$, but we don't consider such a parameter region 
in this paper.) 
\begin{figure}[htb]
\begin{center}
\mbox{\psfig{figure=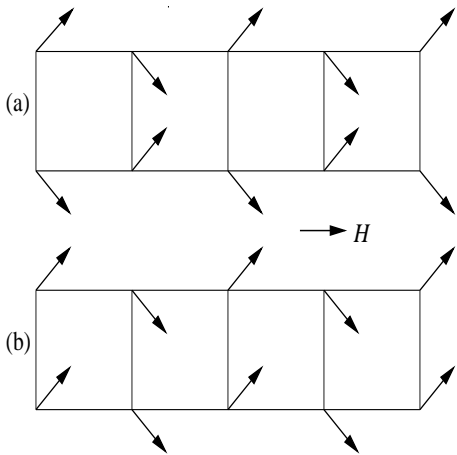,width=6cm,height=6cm,angle=-90}}
\end{center}
\caption{
Canted N\'eel orders of the classical system under external field $H$ 
(a) for $J_2 <1/2$ and (b) for $J_2 > 1/2$. 
}
\label{fig2}
\end{figure}

Even in the magnetic state under external field the two phases  
still can be identified by the different canted N\'eel orders 
shown in the Figs. 2(a) and (b), respectively, in the classical system.
The phase boundary is the same as the nonmagnetic ground state. 
The quantum system is gapless for $0< m < 1$ and it might be
difficult to distinguish the two phases by the dimer picture. 
In this case the classical picture is useful because the gapless phase
is characterized by the power-law decay of the dominant spin correlation
function corresponding to the classical order. 
Thus the quantum system should also have two phases 
like the classical limit. 
The same analysis as the nonmagnetic state indicated the first-order 
boundary for finite $m$. 
We show the boundaries for $m$=1/6, 1/3, 1/2, 2/3 and 5/6 ($L$=12) 
in Fig. 1. 
They exhibit a small $m$ dependence, although it is not so large 
that a field-induced transition between the two phases is 
expected to occur in any realistic situations. 
As the magnetization increases, the boundary tends to approach 
to the classical limit for most magnetizations. 
For $m=1/2$, however, the boundary exhibits a quite different 
behavior and it is close to the nonmagnetic one. 
It implies that the quantum fluctuation is enhanced by the 
frustration particularly at $m=1/2$. 
Thus we consider the possibility of another spin gap 
induced by external field, that is observed as a plateau 
in the magnetization curve at $m=1/2$. 
\begin{figure}[htb]
\begin{center}
\mbox{\psfig{figure=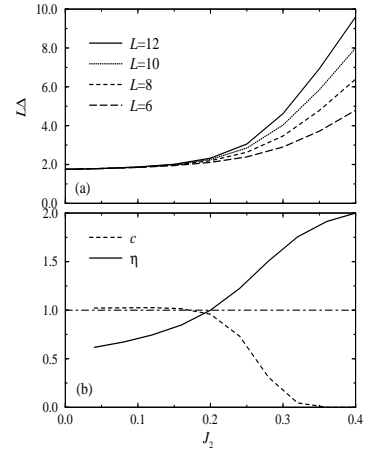,width=6cm,height=6cm,angle=-90}}
\end{center}
\caption{
(a)Scaled plateau $L\Delta$ for $J_1=0.4$,  
(b)central charge $c$ and critical exponent $\eta$, 
plotted versus the NNN coupling $J_2$.  
}
\label{fig3}
\end{figure}

\section{Magnetization plateau}
We consider the magnetization plateau at $m=1/2$. 
The plateau length 
$\Delta \equiv E(M+1)+E(M-1)-2E(M)$ 
is one of useful order parameters\cite{sakai} to investigate the boundary 
between the gapless and plateau phases. 
Since $\Delta$ is the low-lying energy gap, it should obey the 
relation $\Delta \sim 1/L$ in the gapless phase.  
The scaled plateau $L\Delta$ for several $L$ is plotted versus $J_2$ 
with fixed $J_1$ to 0.4 in Fig. 3 (a). 
It suggested that a gapless-gapful transition occurs at $J_2 \sim 0.2$.
To clarify the feature of the transition, we investigate the central charge
$c$ of the conformal field theory (CFT)\cite{cft} 
and the critical exponent $\eta$.  
$\eta$ is 
defined by the asymptotic behavior of the spin correlation function 
$\langle S^+_0S^-_r \rangle \sim (-1)^r r^{-\eta}$.
CFT enables us to estimate $c$ and $\eta$ 
from the low-lying energy spectra of finite clusters,  
using the forms 
$E(M)/L \sim \epsilon (m) -\pi cv_s /6L^2$ and 
$\Delta  \sim \pi v_s \eta / L$ $(L\rightarrow \infty)$, 
where $v_s$ is the sound velocity which is the gradient of the
dispersion curve at the origin.
After some extrapolation to the infinite length limit, 
we show the results of $c$ and $\eta$ for $J_1=0.4$ in Fig. 3(b). 
It justifies that the phase boundary is of the Kosterlitz-Thouless
(KT) transition\cite{kt} with $c=1$ in the gapless phase and $\eta=1$ at the 
critical point. 
Thus we determine the phase boundary as points with $\eta=1$ 
in the $J_1$-$J_2$ plane. 
The result of the KT line 
is shown as solid symbols in Fig. 4 together with the 
first-order boundary indicated as open symbols.
Fig. 4 is a complete phase diagram at $m=1/2$. 
The plateau phase is surrounded by the KT line and first-order line. 
The intersection of the two lines is expected to be a tri-critical
point. 
The present analysis suggested that the plateau appears only in the 
rung-dimer phase.  
The rung-triplet phase reasonably has no plateau,  
because it is equivalent to the uniform $S=1$ chain. 
\begin{figure}[htb]
\begin{center}
\mbox{\psfig{figure=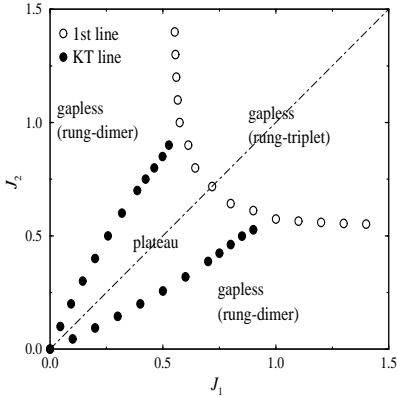,width=6cm,height=6cm,angle=-90}}
\end{center}
\caption{
Phase diagram at $m=1/2$ including the plateau phase. 
The plateau appears only in the rung-dimer phase. 
}
\label{fig4}
\end{figure}

A necessary condition of the presence of the plateau in general 
1D systems was 
rigorously given\cite{oshikawa}
 by $Q(S-m)$. 
$Q$ is the periodicity of the 
ground state and $S$ is the total spin of the unit cell. 
The present case must hold $Q=2$ in the plateau phase at $m=1/2$. 
It suggests that the frustration 
stabilizes the structure where the singlet and triplet rung bonds 
are alternating, 
as is in the case of the zigzag ladder.\cite{tonegawa,totsuka} 

Finally we present the magnetization curves for ($J_1$,$J_2$)=
(0.5,0), (0.5,0.3) and (0.5,0.4) in Fig. 5.  
They were obtained by the size scaling in Ref.\cite{sakai} applied 
to the calculated energy spectra of finite systems up to $L=16$. 
The plateau clearly appears at $m=1/2$ in the latter two cases. 
\begin{figure}[htb]
\begin{center}
\mbox{\psfig{figure=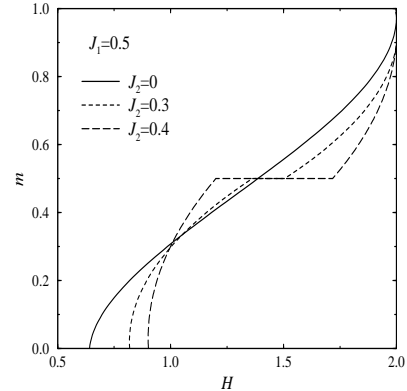,width=6cm,height=6cm,angle=-90}}
\end{center}
\caption{
Magnetization curves for ($J_1$,$J_2$)=(0.5,0), (0.5,0.3) and (0.5,0.4).
}
\label{fig5}
\end{figure}

\section{Summary}
The antiferromagnetic spin ladder with NNN coupling is investigated 
by the exact diagonalization of finite clusters. 
It indicated the existence of the two magnetic phases; 
the rung-dimer and rung-triplet phases, not only for $m=0$ but 
also in the magnetic state. 
It is also found that the magnetization plateau possibly appears 
at $m=1/2$ only in the rung-dimer phase. 


\widetext
\end{document}